# Physical dealloying for two-phase heat transfer applications: pool boiling case

**Artem Nikulin** [a*], **Yaroslav Grosu** [a,b], **Jean-Luc Dauvergne** [a], **Asier Ortuondo** [a], **Elena Palomo del Barrio**[a,c]

[a] Centre for Cooperative Research on Alternative Energies (CIC energiGUNE), Basque Research and Technology Alliance (BRTA), Alava Technology Park, Albert Einstein 48, 01510 Vitoria-Gasteiz, Spain

[b] Institute of Chemistry, University of Silesia in Katowice, Szkolna 9 street, 40-006 Katowice, Poland

[c] Ikerbasque - Basque Foundation for Science, María Díaz Haroko 3, 48013 Bilbao, Spain

**Corresponding author:** *anikulin@cicenergigune.com*

**Abstract**

In this work, physical dealloying was explored as a simple and green method to microstructure the surface of commercial brass for pool boiling heat transfer coefficient enhancement. Three samples were dealloyed for 0.5, 1 and 3 hours at 650 °C, turning the smooth surface into a porous one with a depth of 175, 200 and 223 µm. The boiling experiments carried out in ethanol at 78 °C have shown, that the maximum enhancement of heat transfer coefficient between 110 and 150% was achieved for the sample dealloyed for 0.5 h. Longer intervals of dealloying reduce boiling performance, but it is still much higher compared to smooth brass. This simple method can be customized for various thermal management equipment, such as conventional, plate and micro heat exchangers, all types of heat pipes, HVAC equipment etc., where the heat transfer occurs with phase change.

**Nomenclature**

*Abbreviations*

BSED - backscattered electron detector

EDX - dispersive x-ray spectroscopy

ETD - Everhart-Thornley Detector

HTC – heat transfer coefficient

PD – physical dealloying

SEM – Scanning electron microscope

*Symbols*

a,b – coefficients

$C_f$ – coefficient

Cp – specific heat capacity (J/(kg·K))

F – function

g - acceleration of gravity (m/s$^2$)

h – heat transfer coefficient (W/(m$^2$·K))

$h_{lv}$ – heat of vaporization (J/kg)

k – thermal conductivity (W/(m·K))

P – pressure (Pa)

pr – reduced pressure

Pr – Prandtl number

q″ - Heat flux density (W/(m$^2$))

Ra – mean arithmetic roughness (µm)

T – temperature (°C or K)

k – thermal conductivity (W/(m·K))

l – distance (m)

n - exponent

*Subscripts*

0 – reference value

b – boiling

c – critical

brass – brass

Cu – copper

Cu surf – surface of the copper block

v – vapor

w – wall

*Greeks*

δ – relative difference (%)

µ - dynamic viscosity (Pa·s)

ρ – density (kg/m$^3$)

σ – surface tension (N/m)

# 1 Introduction

Dealloying method is one of the sub-fields of porous metal coatings creation for the two-phase applications. Porous metallic structures have been widely used for heat transfer coefficient (HTC) and critical heat flux enhancement during boiling[1,2]. Different methods were applied to modify heat transferring surface, including physical[3], chemical[4], or combined physico-chemical methods[5]. However, an alloying-dealloying method is currently more frequently found in the literature concerning heat transfer and thermal management, which shows growing interest in this technique among researchers. Tang et al.[5] have reported a significant reduction of wall superheat (about 5 degrees) at the low heat flux region of the boiling curve. In their study, porous coating with a typical pore size of 50-200 nm was created by hot-dip galvanization at the alloying stage and the free corrosion method at the dealloying stage. In the work of Lu et al.[6], a Cu surface was coated by Zn using electroplating. Later the sample was subjected to thermal alloying in a furnace, and finally, free corrosion was used to create pores with the size from 30 to 200 nm. The boiling experiment in water confirmed higher HTC for the porous surface. Huang et al.[7] also applied hot-dip galvanization followed by free corrosion to create a porous copper layer. Performed boiling experiments in water under sub-atmospheric pressure confirmed the better performance of porous surfaces.

Li et al.[8] used the alloying-dealloying method to enhance the boiling performance of sintered Cu matrix. Electroplating followed by free corrosion was used to modify the surface of Cu micro-particles. After treatment, ligaments and pores (less than 150 nm in size) were created on the surface of the particles. Boiling tests in water revealed higher HTC for porous micro-particles in all ranges of tested parameters.

Although efficient in boiling intensification, all alloying-dealloying methods mentioned above use strong acids or bases at the dealloying stage which significantly reduces their ecological efficiency. Physical dealloying (PD) has been recently proposed to fabricate porous metals by utilizing the vapor pressure difference between the alloy components to selectively remove one with a high partial vapor pressure[9]. To the best of our knowledge, up to now, this new dealloying method has never been used for HTC enhancement under pool boiling conditions. In this work, we make a first step towards exploring this chemical-free approach for two-phase heat transfer applications. This method considerably improves ecological efficiency, as it does not imply strong acids and bases required for chemical and physico-chemical dealloying methods, while being a simple and flexible approach. Available publications indicate that by means of physical dealloying the pores from nano[9] to micro size[10] can be created. In terms of metals suitable for physical dealloying, Zn was utilized for porous Co[9] and Cu[10] production. Similarly, by means of physical dealloying, Mg was reported creating porous Nb, Ta, Mo and V[11] and Mn was applied for porous stainless steel synthesis[12]. High temperature and utilization of vacuum are the drawbacks of physical dealloying in comparison with chemical dealloying.

Apart from physical dealloying, other physical methods to create porous layers on metals, like thermal spraying[3] and sintering of particles[13] were reported to be efficient for boiling performance intensification. The advantage of physical dealloying over those methods is the possibility to treat objects of complex shapes. More details on the currently available methods for two-phase heat transfer intensification can be found in the review paper by Liang and Mudawar[14].

## 2 Experimental setup and procedure

### 2.1 Experimental setup

A scheme of the experimental setup is shown in Fig. 1 (a). It is a closed thermosyphon-like system that consists of a condenser, a chiller and a boiler. The main part of the setup is the boiler which is made out of 10 mm stainless steel plates. The width, depth and height of the internal volume of the boiler are 210, 50 and 250 mm respectively. Two plane-parallel sight glasses with 150 mm in diameter viewports serve for the boiling process observation. For more details on the equipment and procedure used to perform the experiments, we refer the reader to our previous work [15].

However, a new heat-transferring block made of oxygen-free Cu was specifically designed to study new flat geometry (Fig. 1(b,c)). In the bottom part of the Cu block, four 60W cartridge heaters are installed. They allow to achieve up to 1.3 MW/m$^2$ of heat flux density. In the upper part of the Cu block (15 mm in diameter), five K-type thermocouples were placed to measure axial temperature distribution. Counting from the top surface, the distances to the thermocouples are 2.5, 5, 10, 15 and 20 mm. However, as shown by the experiments, the lowest thermocouple was located too close to the heat source and was never used in data analyses. To minimize heat losses, the Cu block was placed into the PTFE enclosure (Fig. 1(d). The bottom part of the PTFE enclosure was filled with Loctite 596 silicone to seal and insulate the Cu block from working liquid. Samples for boiling tests (see paragraph 3) were soldered with Sn60Pb40 alloy to the top surface of the Cu block and later sealed with the same Loctite 596 silicone (see final assembly insert in Fig. 1(d)).

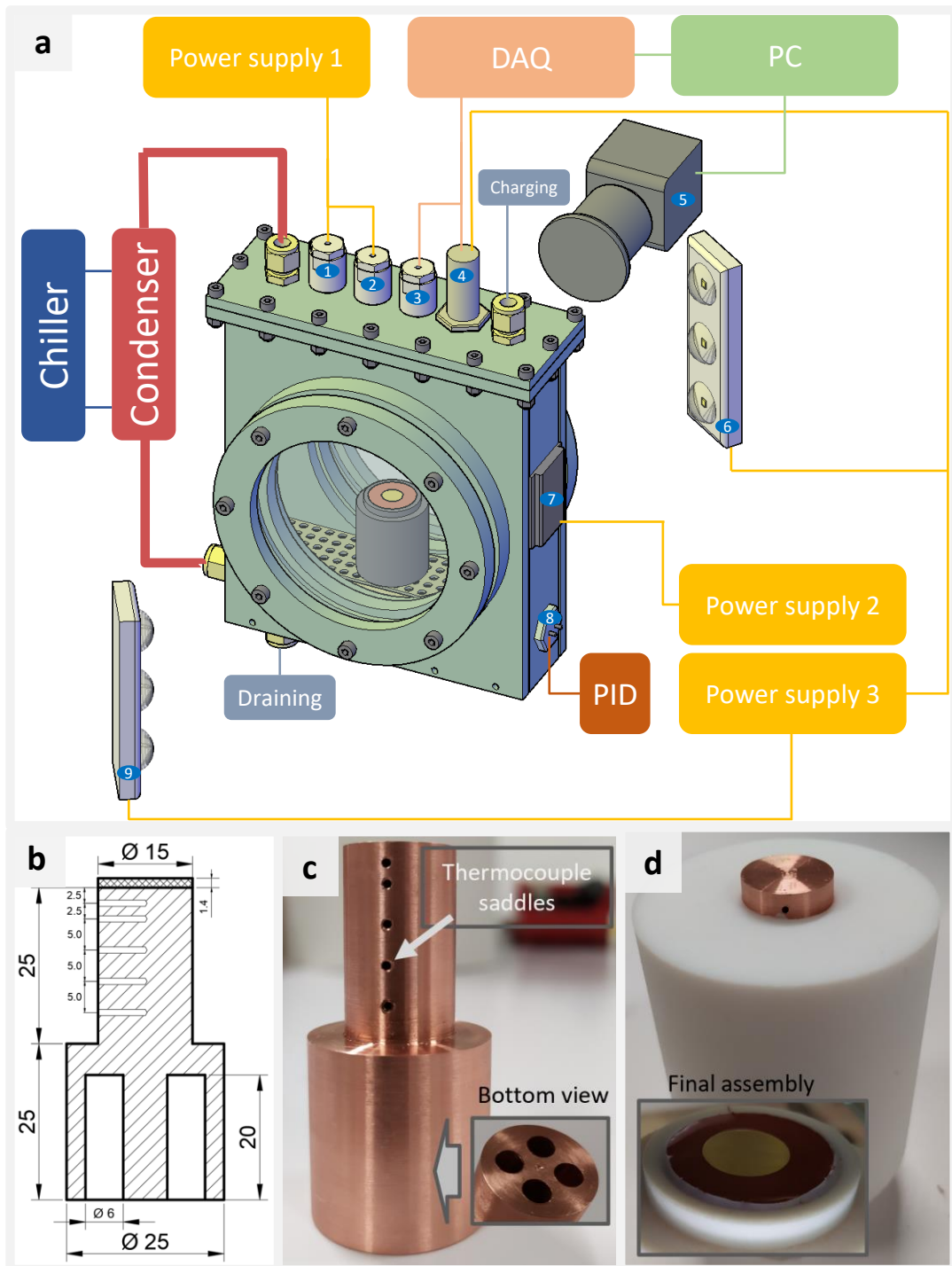

Figure 1. Test rig: (a) scheme (b) Cu block sketch with all dimensions in mm (c) Cu block snapshot (d) Cu block in a PTFE enclosure and final assembly with test surface installed

As a working fluid, 96% ethanol was applied. The working fluid was saturated during the experiment. The saturation temperature of the fluid was 78 °C (± 1 degree). Before the investigation, ethanol was deaerated for 20 minutes by continuous boiling. The experiments were carried out starting with the maximum heat flux, which was later decreased stepwise. Each data point was obtained at a steady state as an average of 300 readings of the Rigol M300 data acquisition system with a sampling rate of 1 Hz. A high-speed camera (i-SPEED 210 from IX-cameras at 800 Hz of the sampling rate) was utilized to image the boiling process.

## 2.2 Data reduction

Assuming steady-state conditions and negligible lateral heat losses, the heat flux density $q''$ dissipated at the test section was calculated using Fourier law for 1D heat conduction

$$q'' = \frac{(T_4 - T_1)k_{Cu}}{l_4 - l_1} \qquad (1)$$

where, $T_1$ and $T_4$ are the temperatures measured by the first and fourth thermocouples installed in the Cu block (K); $k_{Cu}$ is the thermal conductivity of copper (W/(m·K))[16]; $l_1$ and $l_4$ are the distances where the first and fourth thermocouples are placed (m).

The temperature at the top surface of the Cu block was obtained by extrapolation using the data on the lateral temperature distribution. The data measured by four thermocouples installed in the Cu block was fitted by linear equation using Origin 2015 software. Finally, the temperature of the top surface of the Cu block was retrieved from the following equation

$$T = a \cdot l + b \qquad (2)$$

where, $a$ and $b$ are the coefficients obtained after approximation. Considering that $l=0$ at the surface of the copper block, surface temperature $T_{Cu\,surf} = b$.

Wall temperature was calculated knowing the thermal resistance of a sample prepared for boiling tests

$$T_w = T_{Cu\,surf} - \left(q'' \frac{l_{brass}}{k_{brass}}\right) \qquad (3)$$

where, $T_{Cu\,surf}$ is the temperature at the top surface of the Cu block (K); $l_{brass}$ is the thickness of the brass disk (m); $k_{brass}$ is the thermal conductivity of brass (W/(m·K))[17].

Finally, HTC was calculated using the following equation

$$h = \frac{q''}{T_w - T_b} \qquad (4)$$

where, $T_b$ is the boiling temperature measured by a thermocouple immersed in the bath of ethanol (K).

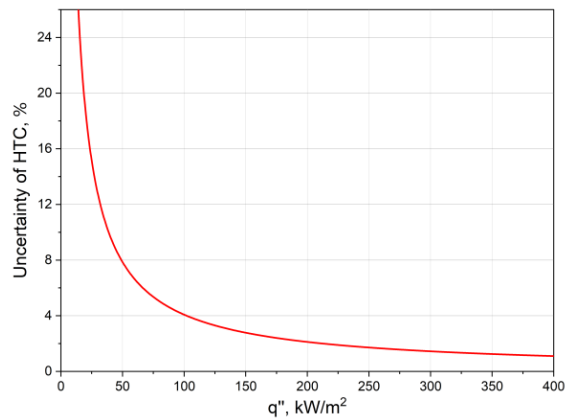

Figure 2. Dependence of relative standard uncertainty of HTC versus heat flux density

All thermocouples used in this study were calibrated to decrease the uncertainty of temperature measurements down to 0.1 K. The uncertainty analysis was carried out according to guidelines for evaluating and expressing the uncertainty[18]. The maximum standard uncertainty of the experimental results was evaluated as 0.2 K for wall superheat and 6300 W/m² or 15% for heat flux density. The dependence of relative standard uncertainty for HTC versus heat flux density is shown in Fig. 2.

**3 Physical dealloying syntheses protocol**

The driving force of PD is the difference in saturated vapor pressure of alloy components at a fixed temperature. According to Sun and Ren[10], Zn pressure is $2.14 \cdot 10^{13}$ times higher than Cu pressure at 500°C. Thus, CuZn alloys, namely brass, can be used for porous Cu manufacturing. In this study, to turn the smooth surface into a porous one, commercially available brass CW614N was utilized. This alloy consists mainly of Cu (57-59%), Zn (37.5-40.5%), and Pb (2.5-3.5%). Some inclusions of Al, Fe, Ni, and Sn are admissible by the CW614N standard[17]. The brass was ordered in the form of a rode. Then, several disks with 15 mm in diameter and 1.4 mm in thickness were lathed out and polished with sandpaper of 1200 grit and thoroughly washed in ethanol. Three pairs of disks were subjected to PD during 0.5, 1, and 3 h, respectively, in a tubular furnace at 650°C, while the absolute pressure was maintained in the range of $0.8\text{-}2.5 \cdot 10^{-2}$ mbar using a fore vacuum pump FB65460 supplied by Welch. Three samples subjected to PD and one brass disk without any treatment were used in boiling experiments. The other three dealloyed samples were used for their characterization.

## 4 Results and discussions

### 4.1 Samples characterization after PD

After dealloying, the brass sample, which is initially of yellow or dull gold color, becomes brownish-orange or copper-like, showing that PD had occurred (see Fig. 3(a)). It is worth mentioning that Zn has low adhesion to the wall of the quartz tube and is easy to remove. Zn deposited on the tube wall and was later extracted in the form of foil is shown in Fig. 3(b). Moreover, dispersive x-ray spectroscopy (EDX) analysis has demonstrated that recovered Zn is pure (Fig. 3(c)) and can be utilized in a new cycle of alloy creation, making this method even more environmentally friendly in comparison with the free corrosion method. EDX for the sample after dealloying (Fig. 3(d)) shows that the upper layer of the porous structure consists of Cu. After one day of storing the oxygen was not detected by EDX.

The structures created on the surface after PD were imaged with the help of a scanning electron microscope (SEM) Quanta 200 FEG equipped with a backscattered electron detector (BSED) and Everhart-Thornley Detector (ETD) in a high vacuum mode at 20 kV. The top planes of created porous surfaces after PD are shown in Fig. 4. As can be seen, the surface becomes covered by ligaments and pores. Clearly, the top edges of ligaments look sharper after 0.5 h of PD, while after 1 and 3 h of PD, they look flatter. The typical size of pores and ligaments ranges from 0.5 to 30 µm. However, nano-roughness was also observed at the surface of ligaments. It is worth mentioning that the surface was structured homogenously, while it is not always the case applying the free corrosion method [4,8]. Moreover, the appearance of the obtained surface structures is similar to those reported by Sun and Ren [19,20] for CuZn alloys, and shows the reliability and reproducibility of the PD method.

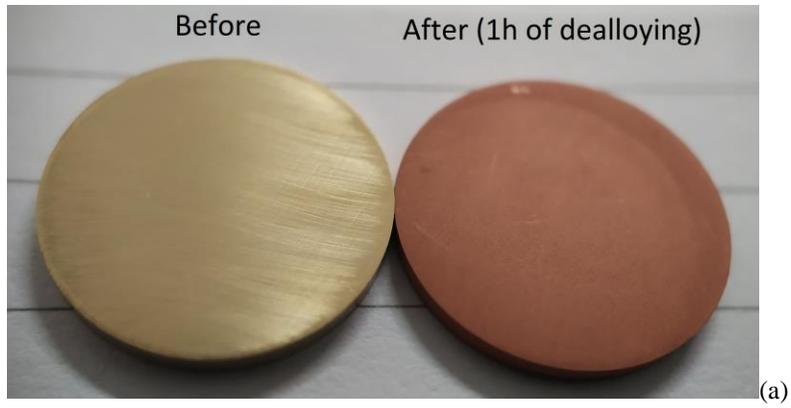

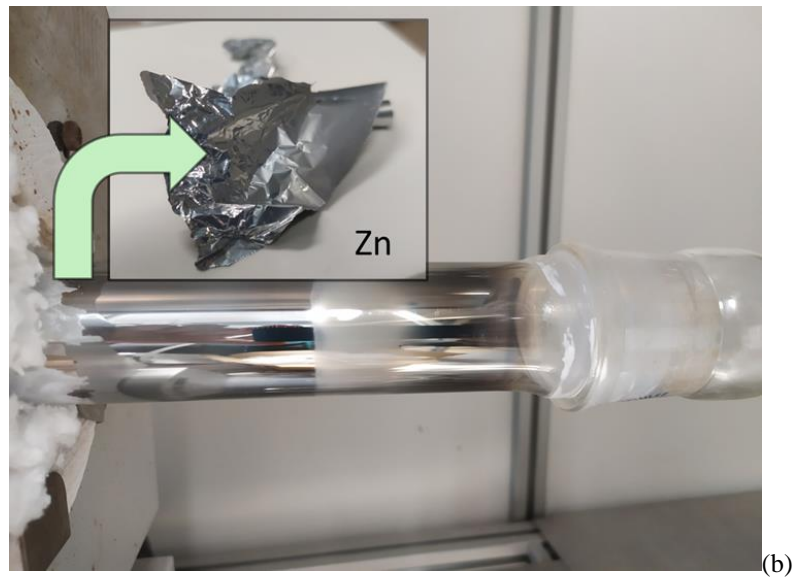

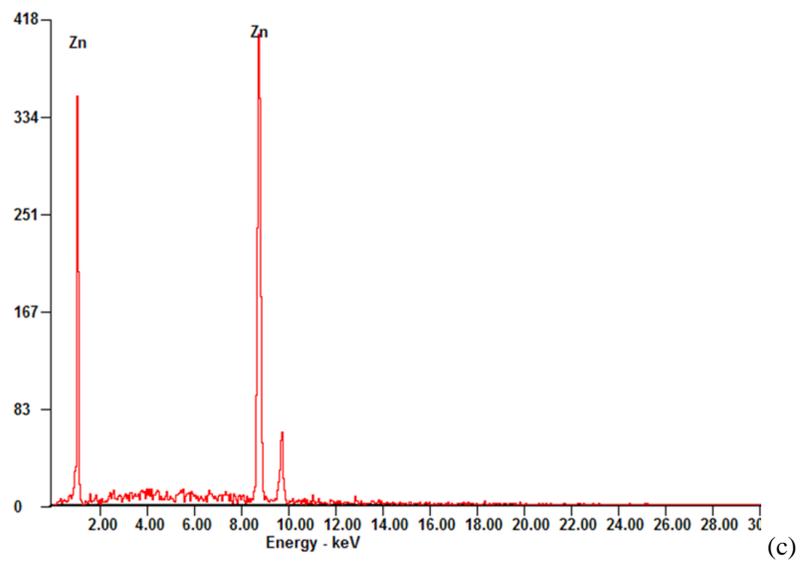

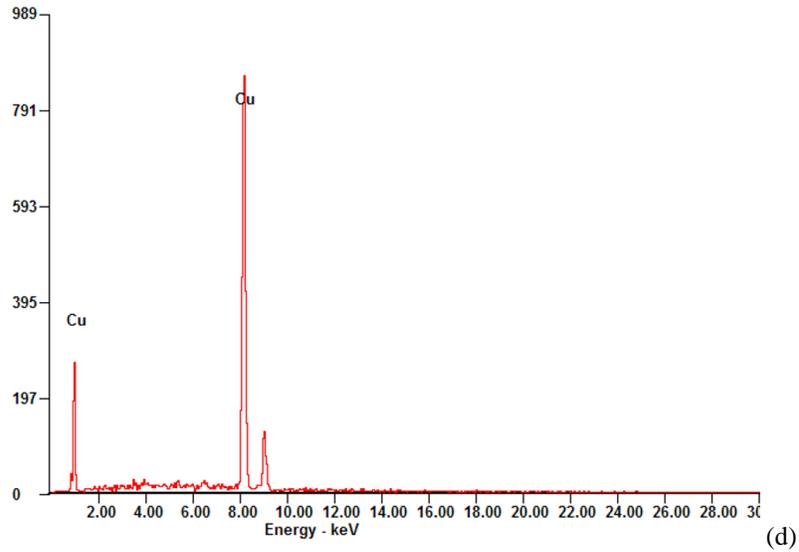

Figure 3. (a) Appearance of the brass samples before and after 1h of dealloying; (b) Zn foil recovered from the quartz tube after dealloying; (c) EDX spectrum of recovered Zn foil; (d) EDX spectrum of the brass sample after 1h of dealloying (taken 1 day after the treatment)

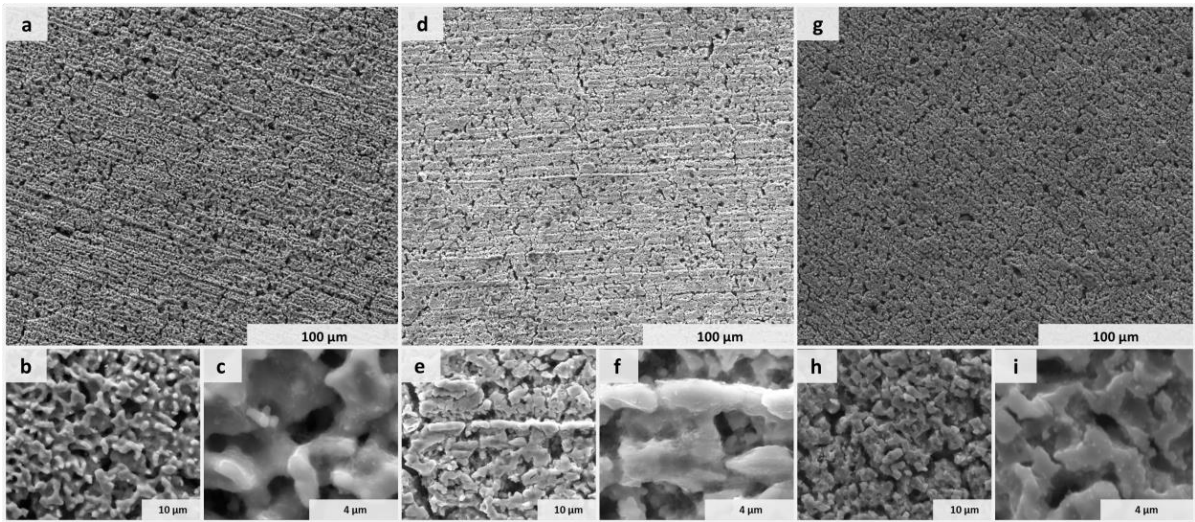

Figure 4. SEM (ETD) images at different magnifications after (a-c) 0.5 h, (d-f) 1 h and (g-i) 3 h of PD

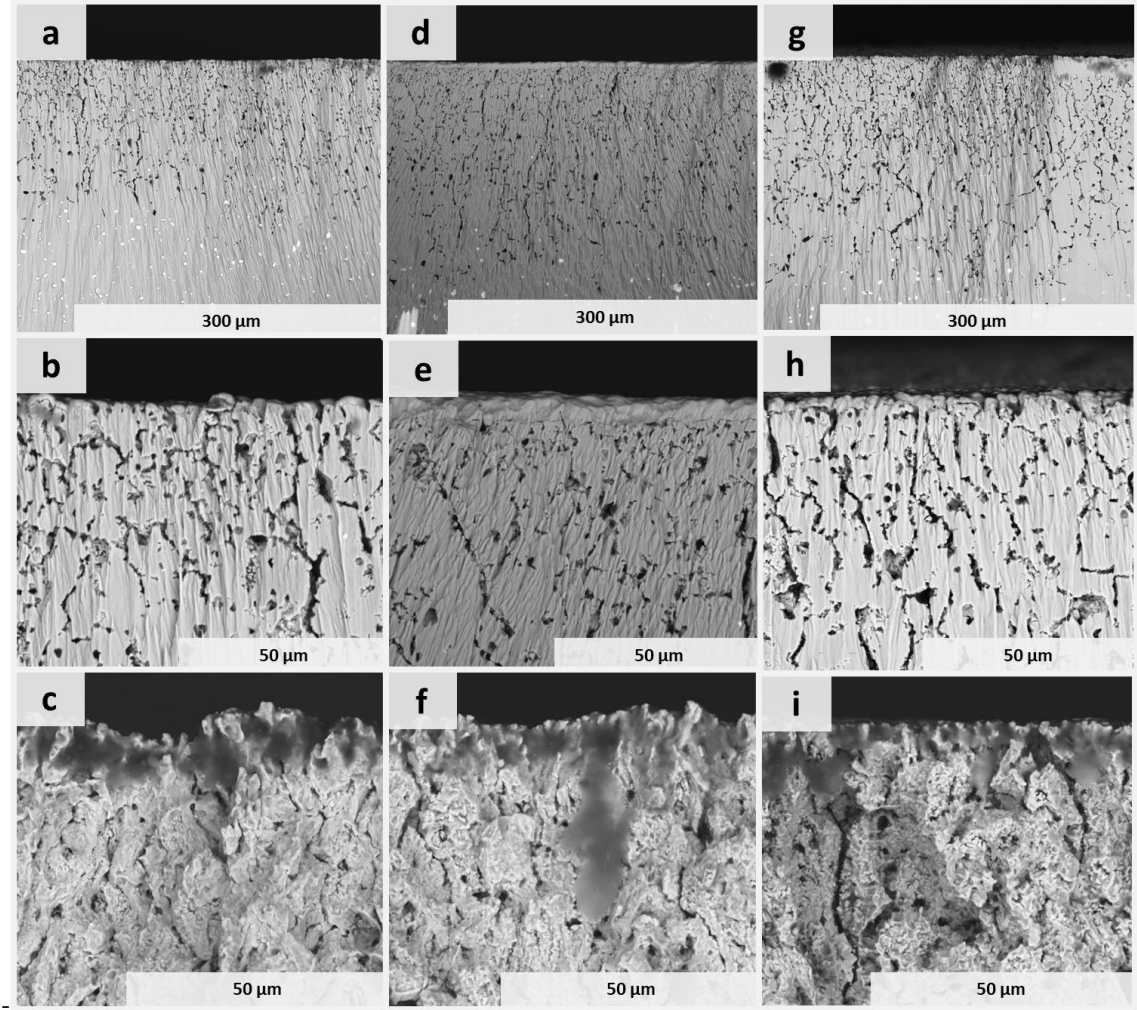

Figure 5. SEM (BSED) images at different magnifications after (a-c) 0.5 h, (d-f) 1 h and (g-i) 3 h of PD. Planes c, f, and i were cross-sectioned by bending until the break; the others were ion milled.

The depth of the microporous layer plays a significant role in the boiling process, the thicker micro[21] and macro[22] porous layers may inhibit HTC. To study the morphology of pores and their depth, the samples were cross-sectioned in two ways: simply bending until the break and using ion milling (Hitachi IM4000Plus). The ion-milled cross-sectional view shown in Fig. 5 revealed that pores are of an irregular shape are formed in the surface of the brass. The width of the pores approximately ranges from 1 to 30 µm. Analysis of the cross-section after the sample was broken shows that the internal surface of pores is very rough and covered by smaller-sized ligaments. The depth of pores depends on the processing time. The average depth measured using SEM images was 175, 200, and 223 µm after 0.5,

1, and 3 h of PD, respectively. Weight loss after PD for each sample was 6.7, 9.2, and 10.0% at corresponding time intervals and with pores depth, demonstrating a severe reduction of PD kinetics in time.

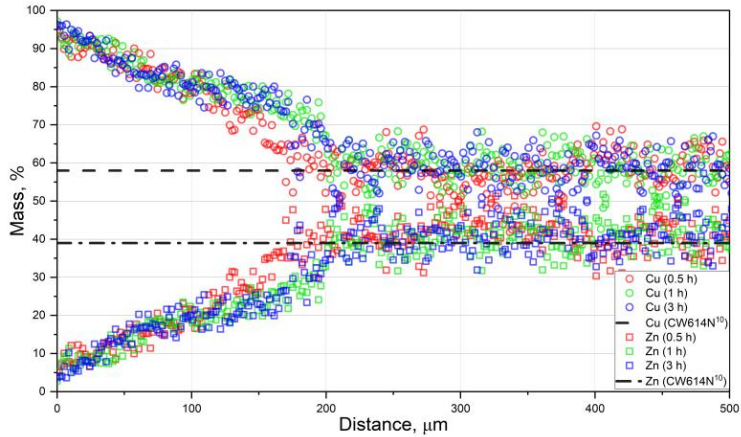

Figure 6. EDX line map of the cross-section samples after PD (SEM images that correspond to these line maps are shown in Fig. 5)

In addition, the composition changes of alloy CW614N caused by PD were quantified using EDX. EDX line mapping of the samples after PD was performed simultaneously with cross-sectional analysis using SEM. According to the results shown in Fig. 6, the content of Zn is reduced to 2.8-5% at the surface of the samples. The composition of Zn and Cu varies almost linearly from the surface to the bulk of the samples, where the components' mass fractions correspond to the CW614N standard[17].

**4.2 New test section validation**

In order to validate the reliability and reproducibility of data obtained with a new test section, several boiling experiments were performed on smooth brass in ethanol. First, the lateral temperature distribution in the heating Cu block was confirmed to be linear. In Fig. 7 (a), one can see that temperature distribution in the Cu block follows linear dependence in the whole range of tested heat flux density with a coefficient of determination $r^2$ equal to or higher than 0.9985. The standard error of Cu block surface temperature $T_{Cu\,surf}$, obtained by extrapolation was always lower than

0.25 K and was considered during the evaluation of the uncertainty of experimental results. The results of heat balance verification in the form of the relative difference $\delta q''$ between $q''$ calculated using equation (1) and $q''$ calculated knowing the electric power supplied to the heat transferring block are shown in Fig. 7 (b). As a result, the relative difference of $q''$ is within the evaluated uncertainty, which confirms the accuracy of the method.

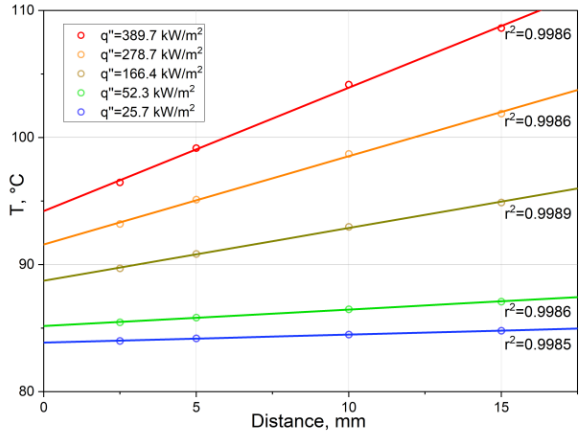

a)

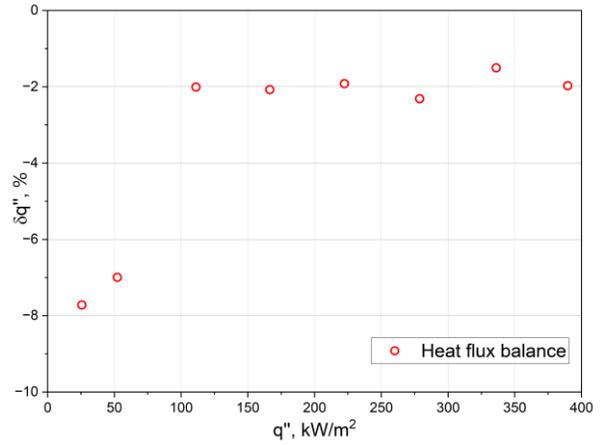

b)

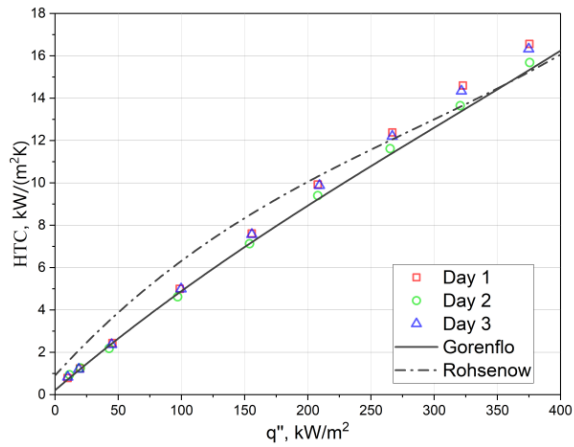

c)

Figure 7. (a) Temperature distribution in a Cu block at various $q''$; (b) accuracy of heat flux measurements; (c) reproducibility of HTC data and comparison with Gorenflo and Rohsenow correlations.

In addition, three boiling curves obtained with one-day intervals are shown in Fig. 7 (c). Deviations of each data point from the average value do not exceed 10% at low heat flow and are within the experimental uncertainty of HTC. Moreover, experimental values of HTC were compared with Gorenflo[16] and Rohsenow[23] correlations. Gorenflo's correlation is expressed by the three following equations

$$h = h_0 F_p \left(\frac{q''}{q''_0}\right)^n \left(\frac{Ra}{Ra_0}\right)^{0.133} \tag{4}$$

$$F_p = 0.7 p_r^{0.2} + 4 p_r + \frac{1.4 p_r}{1 - p_r} \tag{5}$$

$$n = 0.95 - 0.3 p_r^{0.3} \tag{6}$$

where, $h_0$=3970, $q_0''$=2000 and $Ra_0$=0.4 are reference coefficients; $Ra$=0.06 µm is the roughness of brass surface polished with 1200 grit sandpaper[24]; $p_r$=P/Pc is the reduced pressure (a ratio of pressure to critical pressure).

Rohsenow's correlation can be written as follow

$$h = \frac{1}{C_f Pr^n} \frac{q'' C_p}{h_{lv}} \left(\frac{\mu h_{lv}}{q''} \sqrt{\frac{g(\rho - \rho_v)}{\sigma}}\right)^{\frac{1}{3}} \tag{7}$$

where, coefficient $C_f$ and exponent $n$ are individual for each liquid/surface combination and have to be determined experimentally. For the most similar case of ethanol boiling on brass with $Ra$=0.47 µm, Pioro[25] determined $C_f$=0.011 and $n$=0.92. $Pr, C_p, h_{lv}, \mu, \rho, \sigma$ are Prandtl number, specific heat capacity (J/(kg K)), heat of vaporization (J/(kg)), dynamic viscosity (Pa s), density (kg/m³) and surface tension (N/m) of liquid respectively; $\rho_v$ is density of vapor (kg/m³).

The comparison of experimental and predicted by selected correlations HTC is given in Fig. 7 (c). Clearly, the slope and the absolute values are better represented by the Gorenflo correlation. However, the correlation of Rohsenow reasonably follows experimental measurements, considering their uncertainty and the accuracy of the model.

**4.3 Effect of PD on pool boiling**

The effect of PD on the pool boiling process of ethanol at 78 °C is shown in Fig. 8. PD causes strong shift of boiling curves towards lower superheat for more than 10 K. Moreover, the first two data points for the superheat at low heat flux are governed by convection when boiling occurs on a smooth brass surface, while for all surfaces after PD convective regime was not observed (Fig. 8 (a)). The maximum enhancement of HTC was for the sample dealloyed for 0.5 hours: up to 110% at $q''=400$ kW/m$^2$ and up to 150% at $q''=50$ kW/m$^2$. The higher HTC values after PD are associated with three main reasons: *(i)* the number of active nucleation sites on the porous surface is much higher (see Fig. 9), *(ii)* the porous surface has a higher wetted surface area, compared to a smooth surface that contributes to increase HTC[26], and *(iii)* the enhanced surface enforces capillary properties, that help to replenish evaporated liquid in the vicinity of the nucleation sites[14]. To demonstrate the capillary flow in the porous structure, an ethanol droplet was deposited on the brass surface and the surface after PD (see Fig. 10).

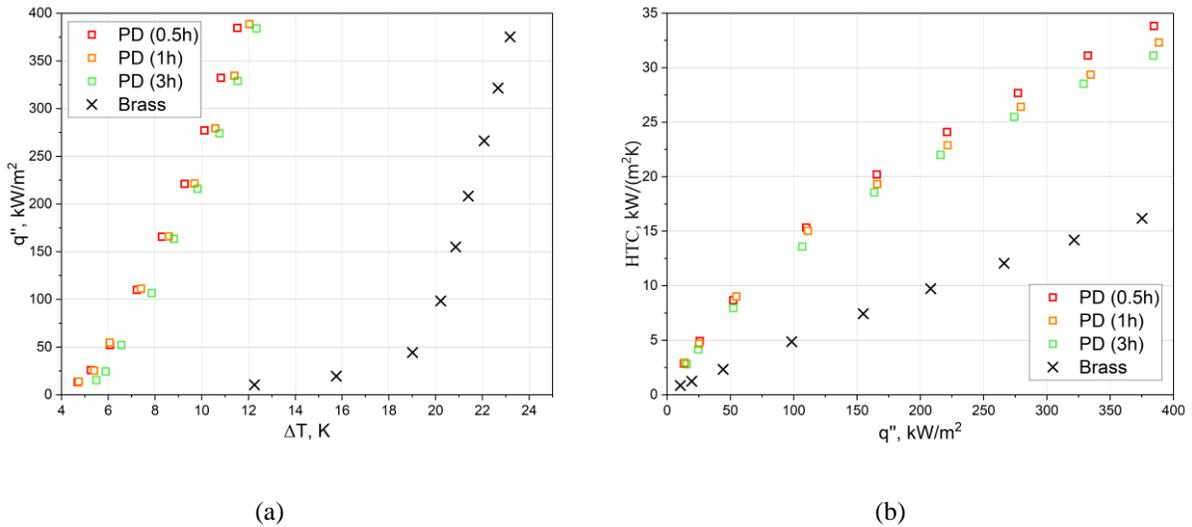

(a)                  (b)

Figure 8. (a) boiling curves and (b) HCT versus heat flux density for ethanol boiling on brass and samples after PD

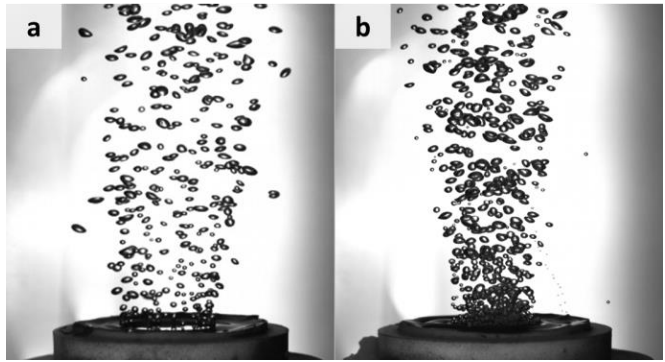

Figure 9. Boiling at $q''=26$ kW/m$^2$ (a) on smooth brass and (b) on the surface after 0.5 h of PD

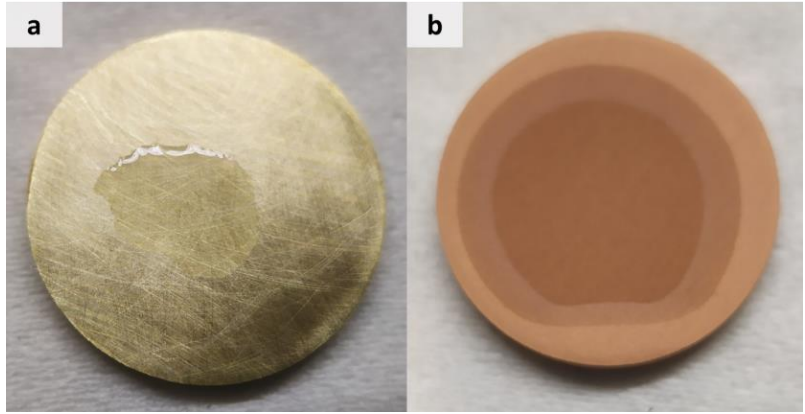

Figure 10. Ethanol droplet appearance on the (a) brass and (b) sample after PD

Finally, it should be noted that the samples dealloyed for a longer time, showed slightly poorer heat transfer performance compared to 0.5 h dealloyed samples. The reduction of HTC indicates that chosen intervals for PD resulted in a shape of ligaments/pores and pores depth that are not optimal. Therefore, optimization of PD intervals is a promising strategy to maximize HTC, which will be explored in the following works.

## 5. Conclusions

This work explores a chemical-free method to create hierarchical porosity on metals´ surface for two-phase heat transfer applications. After vacuum physical dealloying, the surface of commercially available brass was homogenously covered by micropores and ligaments decorated with nanoroughness. The kinetics of PD is highly nonlinear and slows down dramatically in time, which was confirmed both by pores depth measurements and weight loss of the samples. However, in terms of boiling efficiency, only 0.5 hours of PD was sufficient to achieve 100 – 150% enhancement of HTC. The dealloyed Zn was easily and fully recovered in the form of foil after PD process.